\documentclass[preprintnumbers,showkeys,superscriptaddress,showpacs,
nofootinbib,byrevtex,fleqn,prd,notitlepage]{revtex4-2}

\usepackage[utf8]{inputenc}
\usepackage{amsmath,amsfonts,amssymb,amscd,amsxtra,amsthm}
\usepackage{graphicx}
\usepackage{bm,bbold}
\usepackage{multirow}
\usepackage{hyperref}
\usepackage[fleqn]{nccmath}
\usepackage{empheq}
\usepackage{cleveref}
\crefname{equation}{}{}
\input{epsf}
\newcommand{\be}{\begin{eqnarray}}
\newcommand{\ee}{\end{eqnarray}}
\newcommand{\ba}{\begin{array}}
\newcommand{\ea}{\end{array}}

\newcommand{\vb}[1]{\textbf{#1}}

\newcommand{\Slash}[1]{\ooalign{\hfil/\hfil\crcr$#1$}}

\newcommand{\msol}{M_{\mathrm{sol}}}
\usepackage[normalem]{ulem} 
\usepackage[dvipsnames]{xcolor} 
\renewcommand\sout{\bgroup \color{red} \ULdepth=-.5ex \ULset}

\begin{document}
\preprint{    INHA-NTG-10/2022   }
\title{Twist-2 light-quark distribution functions in a singly heavy baryon
in the large $N_c$ limit}
%
\author{Hyeon-Dong Son}
\email[ E-mail: ]{hdson@korea.ac.kr}
\affiliation{Center for Extreme Nuclear Matters (CENuM), 
Korea University, Republic of Korea}
\affiliation{Department of Physics, Inha University, Incheon 22212,
  Republic of Korea} 

\author{ Hyun-Chul Kim}
\email[ E-mail: ]{hchkim@inha.ac.kr}
\affiliation{Department of Physics, Inha University, Incheon 22212,
Republic of Korea}
\affiliation{School of Physics, Korea Institute for Advanced Study
(KIAS), Seoul 02455, Republic of Korea}

\begin{abstract}
A singly heavy baryon can be considered as a state
  consisting of the $N_c-1$ light valence quarks that create the pion mean
  field in the large $N_c$ limit. In the limit of the infinitely heavy
  quark mass, a heavy quark inside a singly heavy baryon is regarded
  as a mere static color source. It is required only to make the
  singly heavy baryon a color singlet. Thus, the $N_c-1$  valence
  quarks govern quark dynamics inside the singly heavy baryon. Within
  this pion mean field framework, we investigate the twist-2
  unpolarized and longitudinally polarized 
  light-quark distribution functions inside charmed and bottom
  baryons at a low renormalization point. We observe that the light
  quarks inside a heavy baryon carry 
  less momentum than those inside a nucleon. This feature is more
  prominent as the heavy quark mass increases. We discuss the
  baryon sum rule, momentum sum rule, and the Bjorken spin sum rule
  for singly heavy baryons. We also discuss the inequality conditions
  for the quark distribution functions.
  In addition, we present the results for the light-quark
  quasi-distribution functions.
\end{abstract}

\maketitle

\section{Introduction}
A singly heavy baryon consists of two light quarks and one heavy
quark. If one takes the limit of the infinitely heavy quark
($M_Q\to\infty$), the spin of the heavy quark is conserved,
which also leads to the spin of the light-quark degrees of freedom is
conserved. This is called the heavy-quark spin
symmetry~\cite{Isgur:1989vq, Georgi:1990um}. In this
limit, the singly heavy baryon is independent of the heavy-quark
flavor, which is called the heavy-quark flavor symmetry. Thus, the
heavy quark inside a singly heavy baryon remains as a mere static
color source. It is only required to make the singly heavy baryon a
color singlet. This indicates that the quark dynamics inside it is
governed by the light quarks. The two light quarks provide the flavor
SU(3) representations of the lowest-lying heavy baryons: the baryon
antitriplet with spin $J=1/2$ and two dengenerate the baryon sextet
with spin $J=1/2$ and $J=3/2$. This degeneracy is removed by the
chromomagnetic hyperfine interaction that arises from the $1/M_Q$
corrections. 

Witten~\cite{Witten:1979kh} proposed that, in the large $N_c$ limit, a
light baryon emerges as a $N_c$ valence quarks bound by the pion mean
field. This picture of the light baryon was realized by the chiral
quark-soliton model ($\chi$QSM)~\cite{Diakonov:1987ty}. The model was
very successful in describing various properties of the nucleon and
low-lying hyperons~(see a review~\cite{Christov:1995vm}). The
$\chi$QSM was further applied to the parton distribution functions
(PDFs) of the nucleon~\cite{Diakonov:1996sr, Diakonov:1997vc,
  Schweitzer:2001sr} and the generalized parton distributions
(GPDs)~\cite{Ossmann:2004bp} (see also a
review~\cite{Goeke:2001tz}). Recently, the $\chi$QSM was 
extended to singly heavy baryons, motivated by
Diakonov~\cite{Diakonov:2010tf}. In the limit of the infinitely heavy
quark mass ($M_Q\to\infty$), a singly heavy baryon can be viewed as a
bound state of the $N_c-1$ valence quarks~\cite{Yang:2016qdz}. A heavy
quark inside the singly heavy baryon remains a mere static color
source. This implies that quark dynamics of the singly heavy baryon is
governed by the light quarks. The presence of the $N_c-1$ valence
quarks create the pion mean field that makes them bound. Combining the
bound state with a heavy quark, we can construct a singly heavy baryon
state. This extended $\chi$QSM has been used to study the charmed
baryon properties such as the mass splitting, isospin mass
splitting, magnetic moments, electromagnetic form factors, radiative
transition form factors, gravitational form factors, axial-vector
transition form factors, and quark spin content~\cite{Kim:2017khv,
  Yang:2018uoj, Yang:2019tst,Yang:2020klp, Kim:2018xlc,Kim:2019rcx,
  Kim:2018nqf,Kim:2019wbg, Kim:2020nug}. 

In the present work, we investigate the light-quark distribution
functions of the singly heavy baryons. Despite the difficulty 
to measure them from hard scattering processes such as deep inelastic
scattering (DIS), the PDFs themselves are fundamental physical
quantities that represent the probability densities to find quarks and gluons
inside a singly heavy baryon with the momentum fraction $x$ and with a 
particular polarization configuration considered. 
Thus, it is of great importance to scrutinize the PDFs so that one can
understand how the heavy and light quarks are distributed inside the
singly heavy baryon. We anticipate that relevant experimental
information on the PDFs inside a singly heavy baryon may be extracted
from fragmentation functions for inclusive heavy baryon productions
with the Drell-Levy-Yan relation used (see Refs.~\cite{Drell:1969jm,
  Pestieau:1969xe, Blumlein:2000wh, Kniehl:2005de, Kniehl:2006mw,
  Goritschnig:2009sq}). 

In Refs. ~\cite{Diakonov:1996sr, Diakonov:1997vc}, the properties
of the twist-2 quark distribution functions in the 
nucleon were studied within the $\chi$QSM. In the present work, we
extend the formalism to the light-quark distribution functions 
inside a singly heavy baryon. We compute the isoscalar unpolarized and
isovector longitudinally-polarized distributions for the light quarks 
and antiquarks inside the singly heavy baryon. We will discuss the
baryon number, momentum and spin sum rules for the light-quark
distributions within the current theoretical framework. We also
discuss the positivity and inequality conditions for the PDFs. We
demonstrate how the $x$ dependence of the ligh quarks and antiquarks
inside the singly heavy baryon is drastically changed, compared with
those inside a nucleon.  We want to mention that in
Ref.~\cite{Guo:2001wi} the behavior of the heavy-quark distribution
was examined in a heavy-quark light-diquark approach. So far as we
know, we show for the first time the light quark distributions inside
a singly heavy baryon. 

This work is organized as follows: In Section~\ref{sec:2}, we present 
the general formalism of the $\chi$QSM. In the following Section, we
show the expressions for the quark and antiquark distribution functions
in a singly heavy baryon and discuss the sum rules.
Then, the numerical results are shown and the positivity and
inequality are discussed. In the penultimate section, the quark
quasi-distributions are considered.  The final section is devoted to
summary and conclusions. 

\section{Baryons in the Chiral quark-soliton model}
\label{sec:2}
In this section, we will briefly explain how the nucleon and singly
heavy baryon emerge respectively as the $N_c$ and
$N_c-1$ valence-quark states bound by the pion mean field. We first
recapitulate the $\chi$QSM for the nucleon and then extend it to the
singly heavy baryon. 

\subsection{Nucleon}
We start from the low-energy QCD partition function given by 
in the large $N_c$ limit: 
\begin{align}\label{eq:partition}
    Z[\psi^\dagger, \psi, \pi^a] = \int \mathcal{D}\pi^a 
    \mathcal{D}\psi\mathcal{D}\psi^\dagger
    \exp \left[\int d^4x \psi^\dagger (i\Slash{\partial} 
    + i M \exp(i \pi^a \tau^a \gamma^5)) 
    \psi \right],
\end{align}
where $\pi^a$ and $\psi$ denote the pseudo-Nambu-Goldstone (pNG)
fields and quark fields, respectively. $M$ is called the dynamical  
quark mass arising from the spontaenous breakdown of chiral symmetry
(SB$\chi$S). In the instanton liquid model for the QCD vacuum at low 
energy~\cite{Diakonov:1985eg, Diakonov:2002fq}, which realizes a
legitimate mechanism of the SB$\chi$S, the value of $M$ is determined
to be $M \approx 350~$MeV. Originally, $M$ is momentum-dependent and
plays a role of an regulator, which causes complexity in calculating
physical observables. In the current work, we turn off its momentum
dependence and introduce a regularization scheme for quark loops with
the ultraviolet (UV) cutoff $\Lambda$ introduced. This implies
that the quark distribution functions are computed at the
normalization point $\mu\simeq \Lambda$, which is approximately given
by $\Lambda\simeq 600$ MeV. In fact, the normalization point can not
be uniquely determined. $\mu$ is proportional to the inverse of the
average instanton size $\bar \rho \approx 1/3~$fm and a dimensional
parameter that varies with bulk properties of the instanton
medium~~\cite{Diakonov:1995qy, Kim:1995bq}. Since $\mu$ is insensitive
to this parameter, we take $\mu=\Lambda\simeq 1/\bar{\rho}$ in the
present work. 

Having integrated out the quark fields in Eq.~\eqref{eq:partition},
we derive the one-loop effective chiral action:
\begin{align}\label{eq:action}
 S_{\mathrm{eff}} = - N_c \mathrm{Tr} \ln \left[i\Slash{\partial} +i
  M U^{\gamma_5} \right], 
\end{align}
where $U^{\gamma_5}$ in Eq.~\eqref{eq:action} is defined as
\begin{align}
    U^{\gamma_5}(x) := \frac{1+\gamma_5}{2}U(x)
    +\frac{1-\gamma_5}{2}U(x)^\dagger, \quad U(x) = \exp(i
  \pi^a(x) \tau^a). 
\end{align}
The pion mean field is defined by a solution of the classical equation
of motion that can be derived from the effective chiral action, which
is expressed by $U_{\mathrm{cl}}(\bm{r})$. The pion mean field
minimizes the nucleon mass, once the symmetries of 
the pion field are all known. Since the pion field carries isospin
indices, one needs to couple them to the spatial components. A minimal
way of this coupling can be done by the hedgehog ansatz
for the pion mean-field:
\begin{align}\label{eq:hedgehog}
   \pi^a(x) \tau^a = \hat n ^a P(r) \tau^a, 
\end{align}
where $\tau^a$ denote the SU(2) Pauli matrices and $n^a$ stands for a radial
unit vector. $P(r)$ is called the profile function of the classical
solution. 

The procedure for deriving the classical nucleon mass is performed by
the self-consistent Hartree approximation. Given a trial profile
function, one diagonalizes the Dirac Hamiltonian in the basis of
$n=(K,P)$ \begin{align} \label{eq:dirac_eq}
    h(U)\Phi_n(\vec x) = E_n \Phi_n(\vec x)
\end{align}
to find the energy spectrum of the quarks in the trial pion mean
field. $K$ and $P$ designate respectively the grand spin defined by
$\bm{K} = \bm{J}+ \bm{T}$ and parity. The eigenvalues and eigenvectors
of the quarks yield a new profile function. We continue this procedure
till we obtain the classical nucleon mass with the minimized energy
functionals:   
\begin{align}\label{eq:MN}
    M_{\mathrm{cl}}  = N_c E_{\mathrm{level}} [U_{\mathrm{cl}}(\bm{r})] 
    + \sum_{n<0} E_{n}[U_{\mathrm{cl}}(\bm{r})] - \sum_{n<0} E_{n}(U=1).
\end{align}
The first term arises from the $N_c$ quarks at the distict level,
whereas the second one emerges from the accumulated energy due to the
polarized vacuum of the Dirac continuum. The last term is required to
subtract the vacuum energy. We need to regularize the second term,
since it comes from the vacuum polarization (quark loops), which
diverges logarithmically. In the current study, we use the
Pauli-Villars regularization scheme with single subtraction to tame
the logarithmic divergence of the quark loops. Also, note that we use
value of $M=420~$MeV in the numerical calculation as used in
describing properties of the light baryons~\cite{Christov:1995vm}. 

The cutoff $\Lambda$ is fixed by computing
the pion decay constant $f_\pi=93$ MeV of which the expression can
also be derived by the effective chiral action~\eqref{eq:action}. The
classical nucleon mass is numerically obtained as 
\begin{align}
M_{\mathrm{cl}}=E_{\mathrm{level}} + E_{\mathrm{cont.}}=1069~\mathrm{MeV},  
\label{eq:2.7}
\end{align}
where $E_{\mathrm{level}}=351$ MeV and $E_{\mathrm{cont.}}=718$ MeV.
While we perform the integral over $\pi^a$ by employing the
saddle-point approximation in the large $N_c$ limit, we have to treat
the zero modes exactly. A rotated field in both three-dimensional
ordinary space and isospin space provides the same nucleon mass,
because the nucleon mass is minimized in both spaces. Thus, we have to
integrate over the rotational and translational zero modes, which is
known as the zero-mode collective quantization. Once we quantize the
classical solution or the chiral soliton, we restore the quantum
numbers of the nucleon, i.e., the nucleon spin, isospin, and
momentum. For details, we refer to Ref.~\cite{Christov:1995vm}.

\subsection{Singly heavy baryon}
A singly heavy baryon can be constructed in exactly the same manner as
we have shown in the previous subsection. The only difference comes
from the fact that the singly heavy baryon consists of $N_c-1$
valence quarks. In the large $N_c$ limit, we can strip off the heavy
quark from the singly heavy baryon. Then the $N_c-1$ valence quarks
will create the pion mean field, which turns out weaker than the
nucleon case. The chiral soliton that constitutes the $N_c-1$ valence
quarks carry the color quantum number. Having performed the
self-consistent method described previously, we obtain the classical
mass for the singly heavy baryon
\begin{align}\label{eq:Msol_HB}
   \msol  = (N_c-1) E_{\mathrm{level}} (U_{\mathrm{cl}})
    + \sum_{n<0} E_{n}(U_{\mathrm{cl}}) - \sum_{n<0} E_{n}(U=1).
\end{align}
Comparing this with Eq.~\eqref{eq:MN}, we find that the prefactor in
the first term is different from that in Eq.~\eqref{eq:MN} whereas the
expression for the energy of the Dirac continuum 
is the same. However, the presence of the $N_c-1$ valence quarks yield
a weaker pion mean field, so that we get $E_{\mathrm{level}} =
213~$MeV and $E_{\mathrm{cont.}}=475~$MeV, thus we have 
$\msol=901$ MeV. Compared with those for the nucleon given in
Eq.~\eqref{eq:2.7}, these results indicate that 
the pion mean field becomes weaker in the precense of the $N_c-1$
valence quarks and produce the continuum energy less
than that for the nucleon. 

Since the heavy quark remains as the static color source, 
the total classical mass of the singly heavy baryon should be given as
the sum of the soliton mass in Eq.~\eqref{eq:Msol_HB} and the heavy
quark mass $M_Q$: 
\begin{align}
    \label{eq:Mcl_HB}
    M^Q_{cl} = \msol + M_Q.
\end{align}
In Refs. \cite{Kim:2019rcx,Kim:2020nug}, it was found that the
the \emph{self-consistent} solution for the pion mean field given in 
Eq. \eqref{eq:Msol_HB} inside a singly heavy baryon plays a 
crucial role for the staibility of the system. Had one used a
parametrized one, the singly heavy baryon would not have satisfied the
stability conditions related to the pressure density that comes from
its $D$-term form factor. Thus, it is essential to construct the pion
mean field in this self-consistent approach. 

\section{Quark distribution functions in the large
  \texorpdfstring{$N_c$}{Nc} limit}
In this section, we provide the expressions for the isoscalar unpolarized 
and isovector polarized distributions derived from the $\chi$QSM. 

\subsection{Light quark and antiquark distribution functions}
Following Ref.~\cite{Diakonov:1997vc}, we define the 
quark and antiquark quasi-number densities for the light quarks 
in a singly heavy baryon as: 
\begin{align}
    D(x,P_h) & = \frac{1}{2E_h} \int \frac{d^3k}{(2\pi)^3}
    \delta\left(x- \tfrac{k^3}{P_h}\right) \int d^3x e^{-i \vb{k}
               \cdot \vb{x}} 
    \langle h | \bar \psi_f(-\vb{x}/2,t)\Gamma
               \psi_f(\vb{x}/2,t)|h_v\rangle ,\\ 
    \bar D (x,P_h) &=  \frac{1}{2E_h} \int \frac{d^3k}{(2\pi)^3} 
    \delta\left(x- \tfrac{k^3}{P_h}\right) \int d^3x e^{-i \vb{k}
                     \cdot \vb{x}} 
    \langle h | \mathrm{Tr} [\Gamma \bar \psi_f(-\vb{x}/2,t)
                     \psi_f(\vb{x}/2,t) ]|h_v\rangle,
\end{align}
where the path-ordered exponential is assumed. 
$x$ is the longitudinal momentum fraction of the quarks and antiquarks
to the heavy baryon momentum. $E_h$ and $P_h$ are the
energy and momentum of the baryon moving with the velocity $v$
\begin{align}
    E_h= \frac{E_h}{\sqrt{1-v^2}}, \quad 
    P_h= \frac{M_h v}{\sqrt{1-v^2}}.
\end{align}
The matrix $\Gamma$ depends on the choice of the Dirac and 
isospin structures. For instance, in the case of the singlet quarks
polarized parellel or anti-parallel with the nucleon momentum,
$\Gamma$ is given by
\begin{align}
    \Gamma = \gamma_0 \frac{1 \pm \gamma_5}{2}.
\end{align}
The PDFs on the light one are obtained by taking the Lorentz boost 
$P_h \to \infty$ (see Refs.~\cite{Diakonov:1997vc, Son:2019ghf}). 

The above matrix elements are written in terms of the 
Dirac spectral representation of the chiral quark-soliton model
Eq. \eqref{eq:dirac_eq}. We will only recapitulate the formulae for
the quark distribution functions, since their derivation was
discussed in Ref. \cite{Diakonov:1997vc} in detail. The isoscalar
unpolarized  quark and antiquark distribution functions 
are written as follows ($x\in [0,1]$) 
\begin{align}
    \label{eq:ISU_model}
    u(x) +  d(x) &= 
    (N_c-1) M_h \int \frac{d^3k}{(2\pi)^3}
                   \Phi^\dagger_\mathrm{level}(\vec k) 
    (1 + \gamma^0 \gamma^3)   \Phi_\mathrm{level}(\vec k)
    \delta(k_3 - x M_h + E_{\mathrm{level}}) \cr
    &\quad  + N_c M_h\sum_{E_n <0} \int \frac{d^3k}{(2\pi)^3}
      \Phi^\dagger_n(\vec k)    (1 + \gamma^0 \gamma^3)  \Phi_n(\vec
      k)-(U \to 1), \\ 
    \bar u(x) + \bar d(x) & = - (u(-x)+d(-x)).
\end{align}
Note that the factor $N_c-1$ for the first term (level contribution)
as the singly heavy baryon consists of $N_c-1$ level quarks. The
second term arises from the vacuum polarization (Dirac continuum). For
the antiquark distribution function, we can easily obtain by using the
property $\bar{q}(x) = - q (-x)$. 
Similarly, the isovector polarized distributions are expressed as 
\begin{align}
    \label{eq:IVP_model}
    &    \Delta u(x) - \Delta d(x) = 
    -\frac{1}{6}(2T_3) (N_c-1) M_h\int \frac{d^3k}{(2\pi)^3}
      \Phi^\dagger_\mathrm{level}(\vec k) 
    (1 + \gamma^0 \gamma^3)   \gamma_5 \tau^3 \Phi_\mathrm{level}(\vec
      k) \cr 
    &   \qquad \qquad \qquad\qquad -\frac{1}{6}(2T_3) N_c M_h 
\sum_{E_n <0} \int   \frac{d^3k}{(2\pi)^3} \Phi^\dagger_n(\vec k) 
    (1 + \gamma^0 \gamma^3)   \gamma_5 \tau^3 \Phi_n(\vec k)-(U \to
      1), \cr 
& \Delta \bar u (x) - \Delta \bar d(x) =  \Delta u (-x) - \Delta
  d(-x). 
\end{align}

\subsection{Heavy quark distribution functions}
Concerning the valence charm heavy quark, we only consider it
as a static color source. So, we obtain the matrix element for the
heavy quark PDFs as  
\begin{align}
    \frac{1}{2E_h} \int \frac{d^3k}{(2\pi)^3}
    \delta \left(x- \tfrac{k^3}{P_h}\right) \int d^3x e^{-i \vb{k}
  \cdot \vb{x}} 
    \langle h | \bar Q_f(-\vb{x}/2,t)\Gamma
  Q_f(\vb{x}/2,t)|h_v\rangle. 
\end{align}
In the heavy quark limit $M_Q \to \infty$, its four-velocity is
fixed. Thus, in this limit, one can eliminate the 
trivial kinetic part that depends on the momentum for each 
velocity~\cite{Georgi:1990um, Wise:1993wa}:
\begin{align}
    \Psi_Q(x) = e^{-iM_Q v \cdot x} \tilde \Psi_Q(x).
\end{align}
Treating $\tilde \Psi$ as the non-interacting field, 
one obtains the unpolarized and polarized distributions for the heavy
quark 
\begin{align}
   Q(x) &= \delta(x-M_Q/M_h), 
\label{eq:HQ_isu}\\
    \Delta Q(x) &= -\frac{1}{3} \delta(x-M_Q/M_h) 
\label{eq:HQ_ivp}.
\end{align}
As expected, they are given as $\delta-$functions 
with the fixed point $x = M_Q/M_h$. In Ref.~\cite{Guo:2001wi}, the   
heavy-quark distribution functions in a singly heavy baryon 
were studied within a heavy-quark light-diquark picture,  
where the interactions between the heavy-quark and the diquark in a
heavy baryon were considered. It was observed that the heavy-quark
distributions have a finite size. In general, one expects a similar
behavior once the interaction between the heavy quark and the light 
soliton is considered.  

\subsection{Sum rules}
Let us examine the baryon number and the momentum sum rules
as the leading and next-to-leading Mellin moments of the isoscalar
unpolarized distributions. Using Eqs.~\cref{eq:ISU_model,eq:HQ_isu},
we obtain the following sum rules: 
\begin{align}
   & \int^1_0 \; dx \; (u(x)+d(x) - \bar{u}(x) - \bar{d}(x)) = N_c-1 = 2, 
   \label{eq:bn_sumrule}
   \\ 
   & \int^1_0 \; dx \; Q(x) = 1, \\
   & \int^1_0 \; dx \;x( u(x)+d(x) + \bar{u}(x) + \bar{d}(x)) 
   = \msol / M_h,\label{eq:momentum_sumrule} \\
   & \int^1_0 \; dx \;x Q(x) = M_Q/M_h.
\end{align} 
Note that the baryon number sum rule Eq.~\eqref{eq:bn_sumrule} is
satisfied by the $N_c-1$ light quarks occupying the bound level,
i.e. no contributions from the polarized vacuum. In the picture of the
$\chi$QSM, the momentum of the singly heavy baryon is carried only by
the quarks. Thus, the light-quark momentum sum rule is identified as
the ratio of the soliton mass to the baryon mass as
$M_{cl}/(M_{cl}+M_Q)$. These results are also predicted in a
study of the energy-momentum tensor (EMT) form factors of the heavy
baryons~\cite{Kim:2020nug}, because the momentum sum rule given in
Eq.~\eqref{eq:momentum_sumrule} corresponds to the mass form factor
$A_q(0)$ in Ref.~\cite{Kim:2020nug}. For the sake of comparison, we
provide the sum rules for the nucleon:  
\begin{align}
    & \int^1_0 \; dx \; (u(x)+d(x) - \bar{u}(x) - \bar{d}(x)) = N_c =
      3, \\  
    & \int^1_0 \; dx \;x( u(x)+d(x) + \bar{u}(x) + \bar{d}(x)) = M_q =
      1. 
 \end{align} 

Now let us discuss the Bjorken spin sum rule. The leading moment of 
the isovector polarized quark distribution is proportional to 
the isovector axial charge $g_{A,q}^{(3)}$ as follows 
\begin{align}
 &\int^1_0 \; dx \;\left(\Delta u(x) - \Delta d(x)+\Delta \bar u(x) -
   \Delta \bar d(x)\right) 
  = g^{(3)}_{A,q}. \label{eq:spin_sumrule} 
\end{align}
 We find that the isovector axial charge does not depend
on the heavy baryon mass $M_h$. Also, the singlet heavy quark spin
   is computed in a similar way and we obtain
   \begin{align}
  \Delta Q \equiv  \int^1_0 \; dx \; \Delta Q(x) = -1/3,  
   \end{align}
which is identical with that derived from the nonrelativistic quark
model~\cite{Suh:2022atr}. 
  
\section{Numerical results and discussions}
\label{sc:result}
We adopt the numerical method to compute the quark distribution
functions, which was already explained in
Ref.~\cite{Diakonov:1997vc} in detail. Since we focus on the light
quark distributions inside $\Sigma_c^+$ and $\Sigma_b^+$, we introduce
the heavy quark masses $M_c=1300$~MeV and $M_b=4200$~MeV, which are
close to those given by the PDG~\cite{Workman:2022ynf}. We want to mention that
specific values of the heavy-quark mass are not important. We
are interested in how the quark parton distribution functions change
as $M_Q$ increases, so we will see how the difference between the values
of $M_c$ and $M_b$ come into play. We will show that when $M_Q$
increases, the PDFs undergo drastic changes. 
Since we take the limit of $M_Q\to \infty$, we ignore $1/M_Q$
corrections: the pion mean-field for both heavy baryons is identical
and the differences of the distributions are originated from the
varied $M_Q$ values at the leading order.  

In the current work, we will employ the ansatz for the profile
function that was suggested in Ref.~\cite{Schweitzer:2012hh}
\begin{align}
    \label{eq:arctan_tanh}
    P(r) = 2 \arctan(r_0^2/r^2 \tanh(b M r) )
\end{align}
instead of using the self-consistent one. 
The function \eqref{eq:arctan_tanh} correctly reproduces 
the linear behavior of the self-consistent mean-field profile as
$r\approx 0$. We fit the parameters $r_0$ and $b$ using the
self-consistent profile functions for the $N_c$ and $N_c-1$ valence
quarks. It was already shown that the shapes of the distribution
functions with the ansatz in the interpolation
approximation~\cite{Diakonov:1987ty} vary within only a few percents
in comparison with those obtained by using 
the self-consistent one. Thus, the numerical results presented in the
following sections are obtained with Eq.~\eqref{eq:arctan_tanh}
utilized The contribution of the Dirac continuum was derived by the
interpolation formula~\cite{Diakonov:1987ty}.

\subsection{Isoscalar Unpolarized Distributions}
We first check that the baryon number sum rule 
Eq.~\eqref{eq:bn_sumrule} is well satisfied numerically. However, the  
momentum sum rule Eq. \eqref{eq:momentum_sumrule} is broken by around
$1~\%$, because the momentum sum rule is related to the saddle-point
equation for the pion mean-field. This small discrepancy arises from
the ansatz in Eq.~\eqref{eq:arctan_tanh}. If one uses the
self-consistent profile functyion, the momentum sum rule is exactly
satisfied. 

\begin{figure}[htbp]
    \centering
    \includegraphics[width=4.9cm]{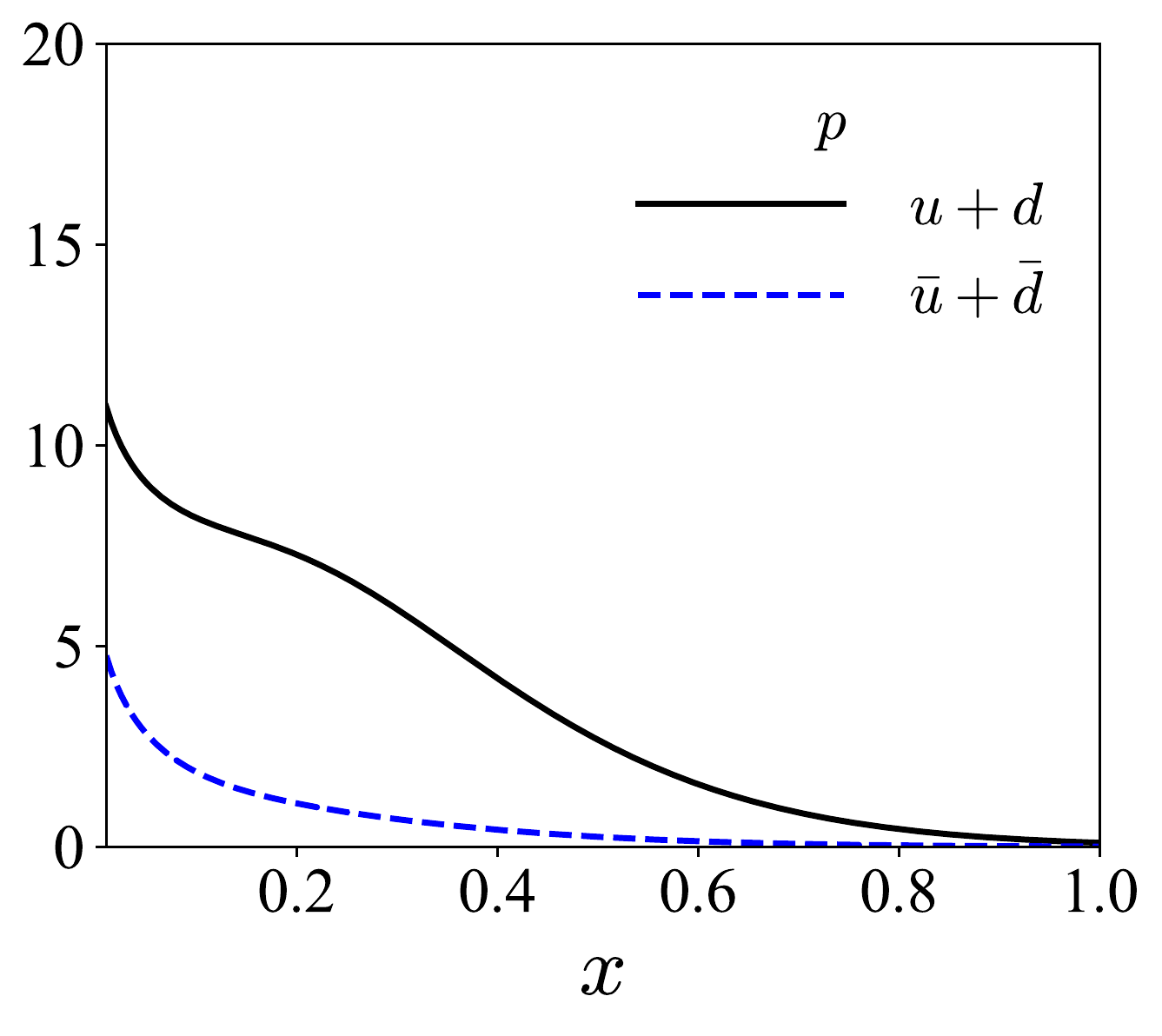}
    \includegraphics[width=4.9cm]{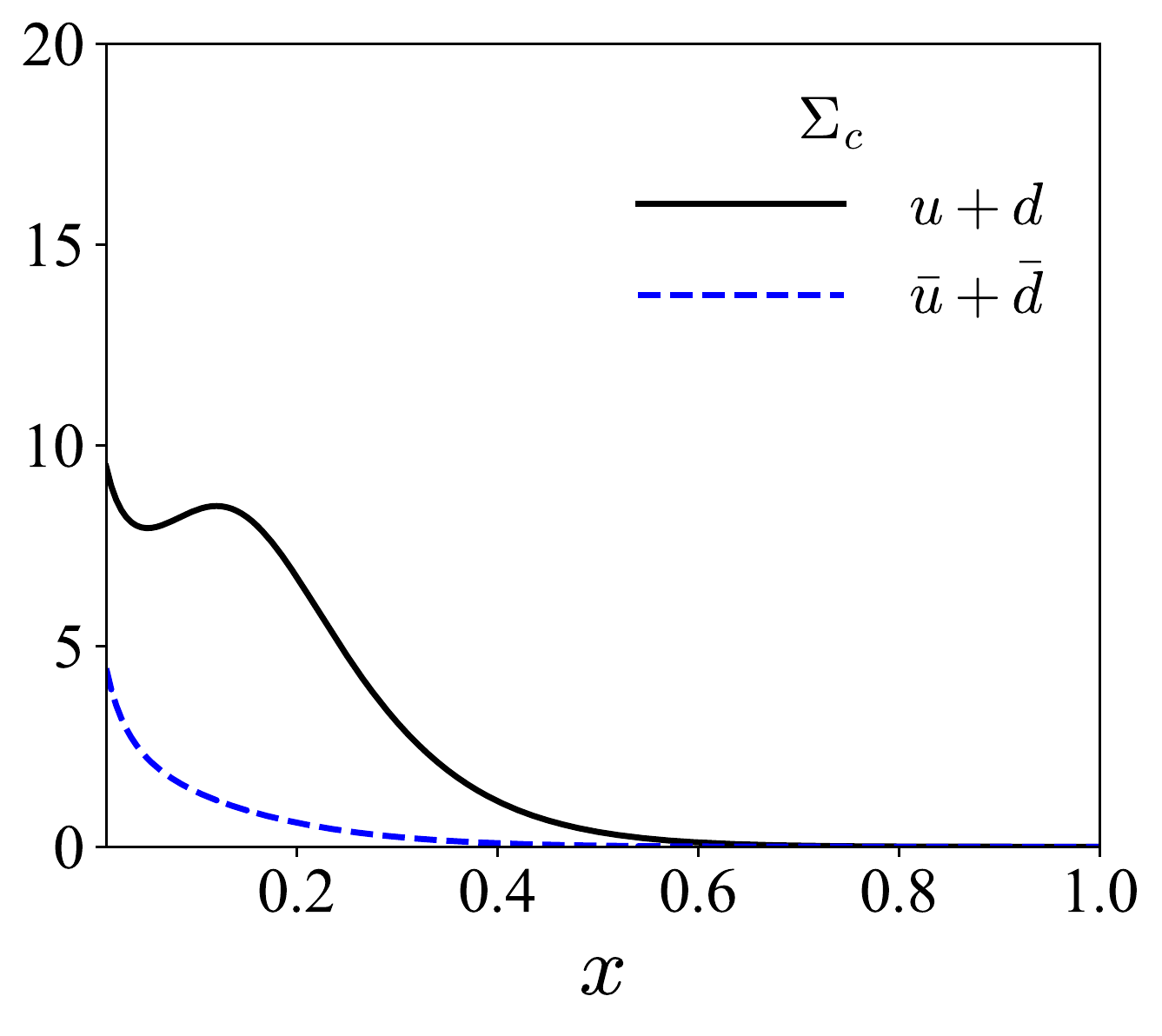}
    \includegraphics[width=4.9cm]{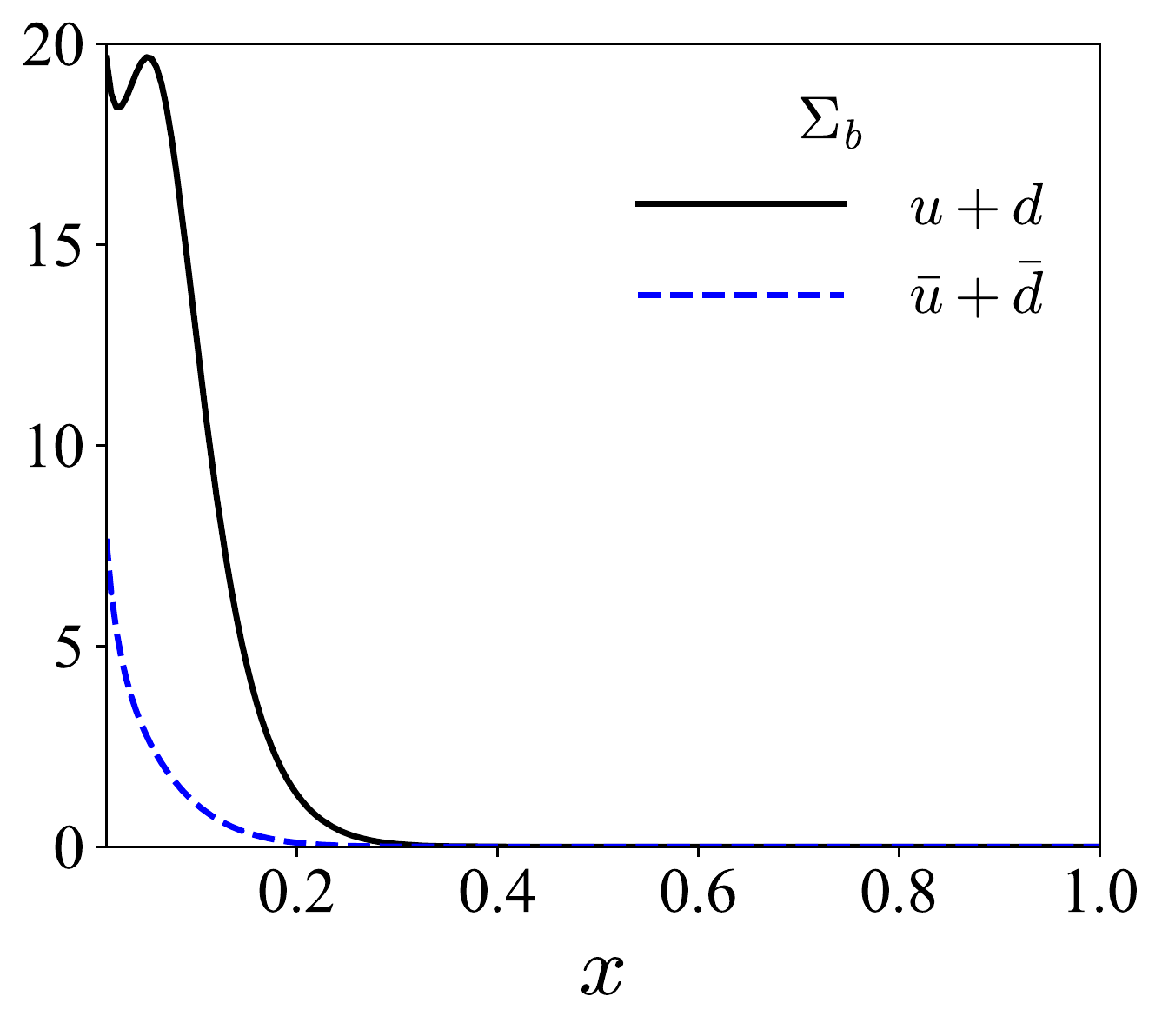}
    \caption{Isoscalar unpolarized quark distribution $u(x)+d(x)$ for quark 
    (black solid curve) and antiquark (blue dashed one) for the
    proton, $\Sigma_c$, and $\Sigma_b$ drawn from the left to the
    right panels, respectively.}
    \label{fig:1}
\end{figure}
In Fig.~\ref{fig:1}, we depict the isoscalar unpolarized quark
distributions $u(x)+d(x)$ in the solid curve as well as the
antiquark distribution $\bar u(x)+ \bar d(x)$ in the dashed one for
the nucleon, $\Sigma_c$, and $\Sigma_b$ from the left to the
right panels, respectively. The light quark and antiquark 
distributions in the $\Sigma_c$, compared with those in the proton,
are more concentrated in the smaller $x$ region. We observe that as
$M_Q$ grows (from $c$ to $b$), the quark and antiquark distributions are 
extremely squeezed to the even smaller $x$ region. We will later
discuss this behavior in detail. 

\subsection{Isovector Polarized Distributions}
Before we proceed to present the numerical results for the isovector
polarized quark distribution functions, we examine the Bjorken spin
sum rule given in Eq. \eqref{eq:spin_sumrule}. We obtain numerically
the axial charge 
\begin{align}
    &    g_{A,q}^{(3)}= \int^\infty_0 \; dx \; 
    (\Delta u(x) - \Delta d(x) - \Delta \bar{u}(x) + \Delta
      \bar{d}(x)) = 0.7   
 \end{align} 
which are identical to the $\Sigma_c$ and $\Sigma_b$ baryons ($T_3=1$).
We compare this numerical value with that obtained from the study of
axial-vector form factors for singly heavy baryons computed within the
$SU_f(3)$ chiral quark-soliton model~\cite{Suh:2022atr}, which was
given as $g_{A,q}^{(3)} = 1.026$. At first glance, this discrepancy
seems to be very large and concerns us. However, we consider in the
current work only the leading contribution in the large $N_c$ limit,
whereas in Ref.~\cite{Suh:2022atr} the rotational $1/N_c$ corrections
are included. Thus, this discrepancy will be removed once the rotational  
$1/N_c$ corrections are introduced to the quark distribution
functions. For comparison, we present the spin sum rule for
the proton ($T_3=1/2$), which is $g_{A}^{(3)}=0.95$. By the same
token, we can get $g_A^{(3)}=1.163$~\cite{Suh:2022atr}, when we
include the rotational $1/N_c$ corrections.  

\begin{figure}[htbp]
    \centering
    \includegraphics[width=4.9cm]{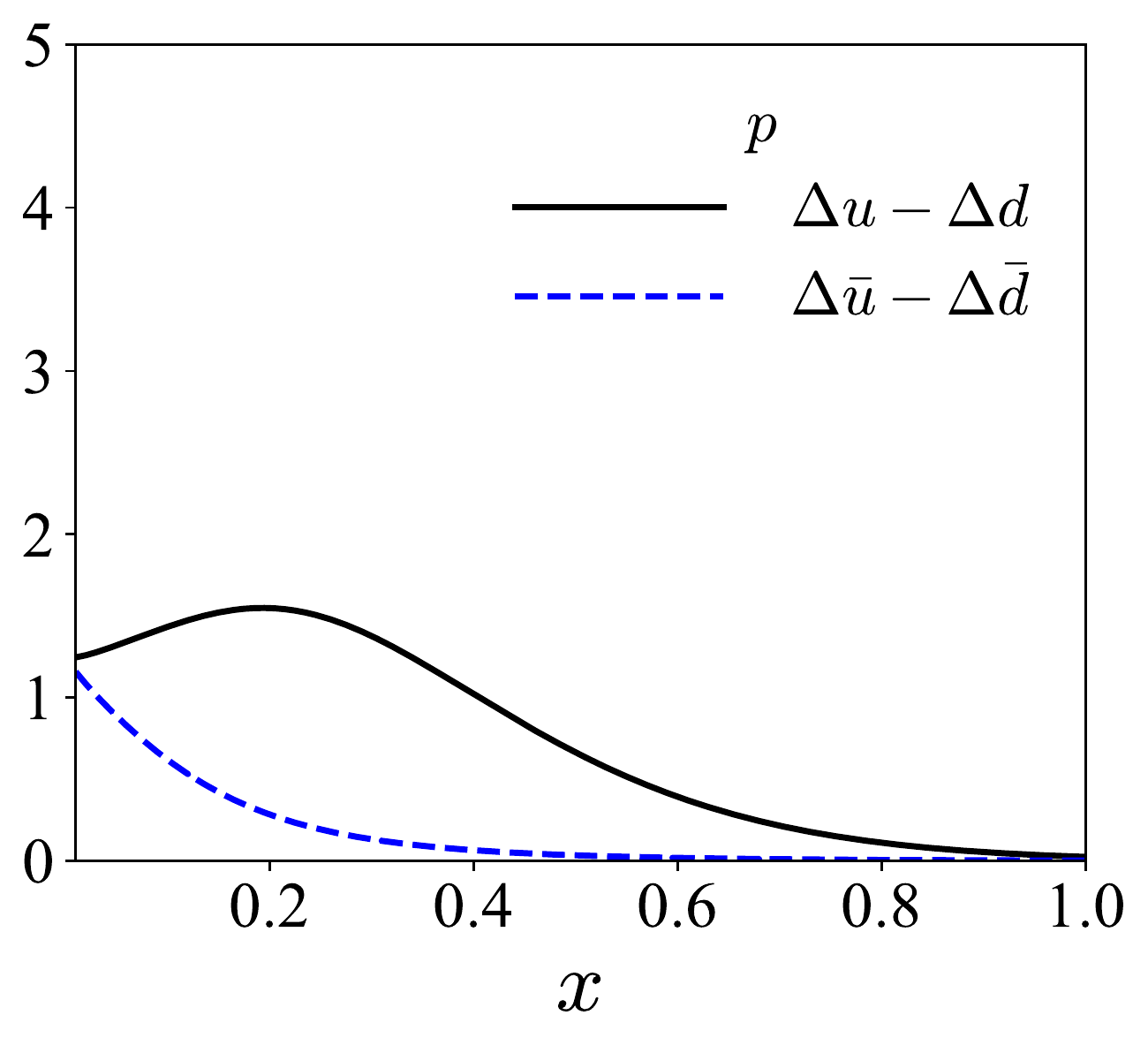}
    \includegraphics[width=4.9cm]{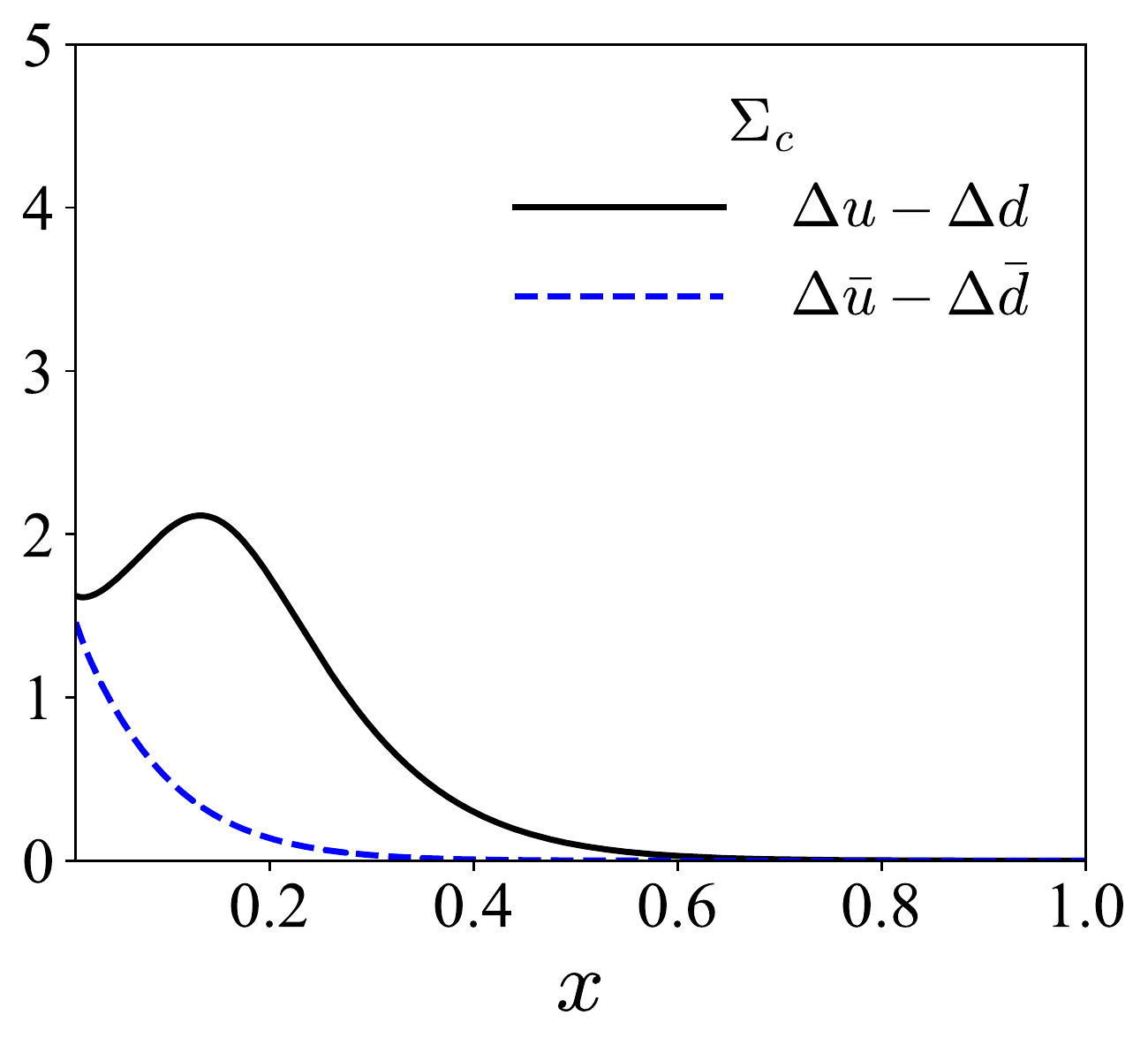}
    \includegraphics[width=4.9cm]{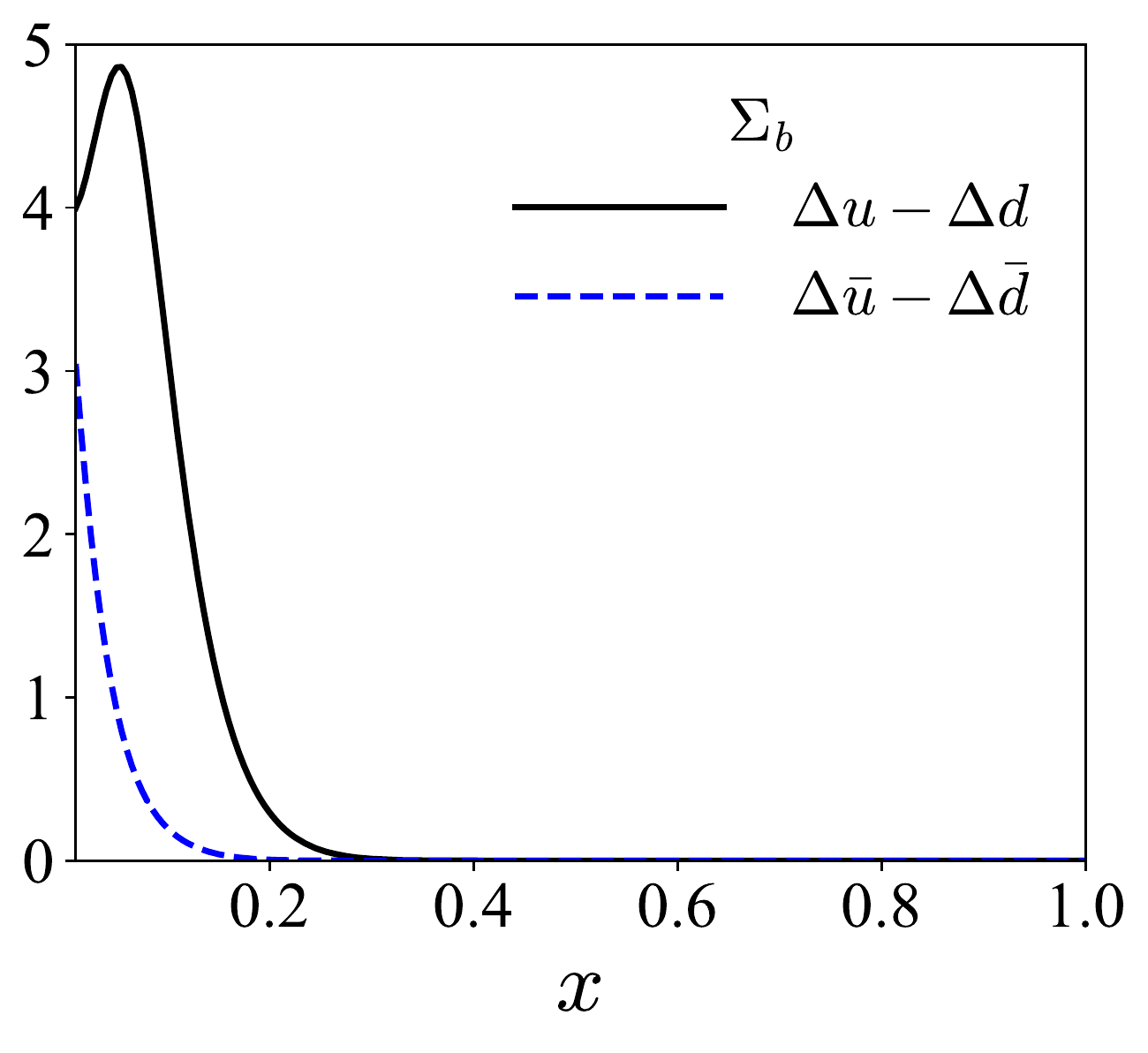}
    \caption{Isovector polarized quark distribution $\Delta u(x) -
      \Delta d(x)$ for the quark (black solid curve) and antiquark
      (blue dashed one) for the proton, $\Sigma_c^{++}$, and
      $\Sigma_b^{++}$ drawn from the left to the
    right panels, respectively.
}
    \label{fig:2}
\end{figure}
In Fig.~\ref{fig:2}, the isovector polarized quark (solid curve) and 
antiquark (dashed one) distributions are displayed for the proton, 
$\Sigma_c^{++}$, and $\Sigma_b^{++}$ (isospin $T_3=+1$) consecutively
from the left to the right panels. We find the similar
$x$-dependence of the quark and antiquark distributions in a singly
heavy baryon.  

 \subsection{Momentum distribution of quarks in a heavy bayron}
In the infinitely heavy-quark mass limit, we expect that the heavy
quark carries the entire longitudinal-momentum of the system. 
When one uses a large but finite heavy-quark mass, the light quarks start to share the
momentum. As already shown in Fig.~\ref{fig:1}, the light quarks 
inside a heavy quark carry a much smaller portion of the momentum of
the singly heavy baryon. That is, the light-quark distribution
functions exhibit the shapes squeezed to the smaller $x$ region. 
When $M_Q$ increases from $M_c$ to $M_b$, the shape is further
squeezed to the even smaller $x$ region. We observe a similar tendency
in the isovector polarized quark distribution functions as shown in
Fig.~\ref{fig:2}. 

To analyze this behavior quantitatively, we define the following
integral for the isoscalar unpolarized distributions 
\begin{align}\label{eq:sum_int_I}
  \mathcal{I}_h(y) := M_h \int^y_0 \;dx\;x
  (u(x)+d(x)+\bar{u}(x)+\bar{d}(x)), 
\end{align}
where $h$ denotes the baryon of interest: the proton, $\Sigma_c$,
 or $\Sigma_b$. $y$ stands for the upper limit of the integral, $y\in
 [0,1]$. 
\begin{figure}[htbp]
    \centering
    \makebox[\textwidth]{
    \includegraphics[width=6.5cm]{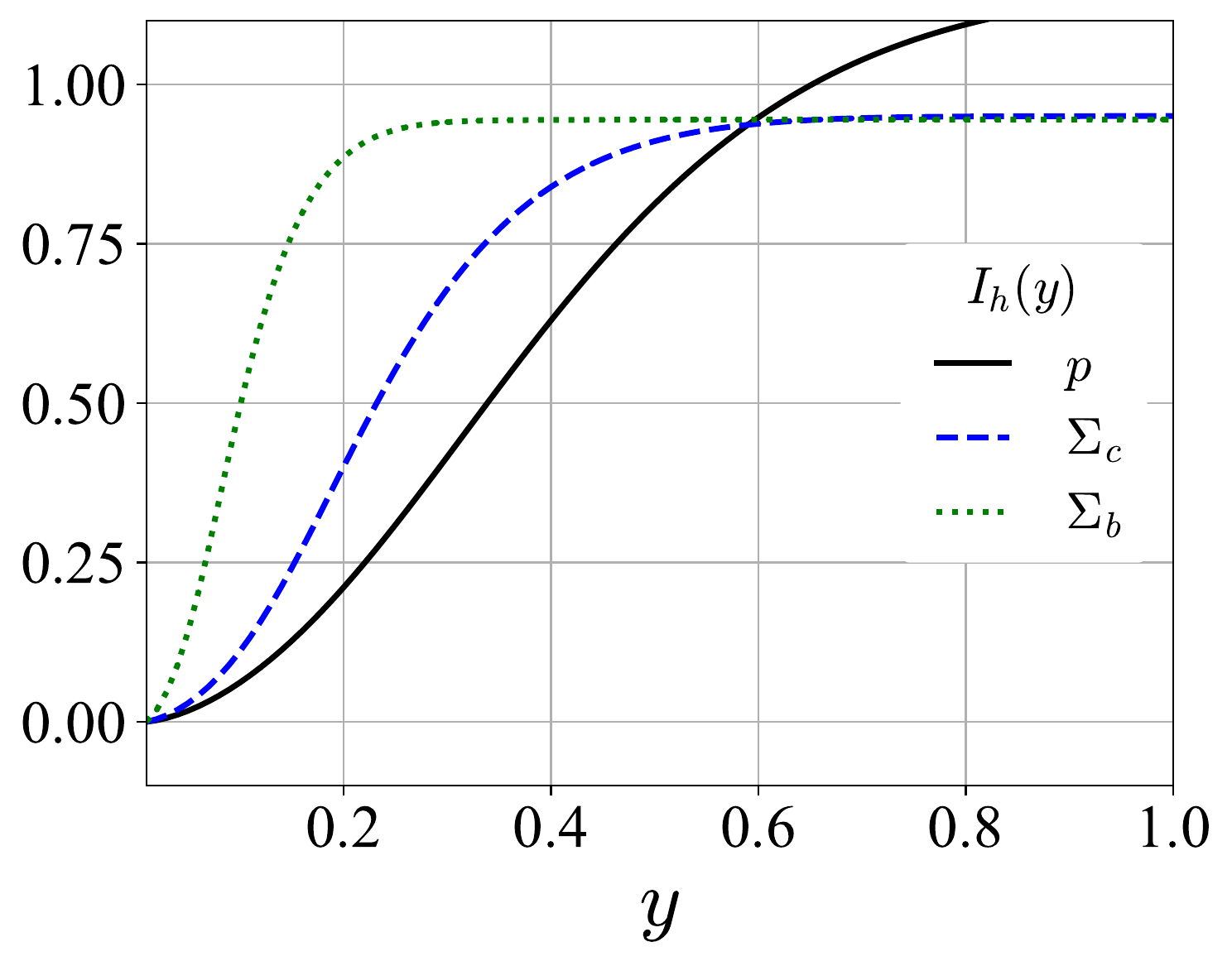}
    \includegraphics[width=6.5cm]{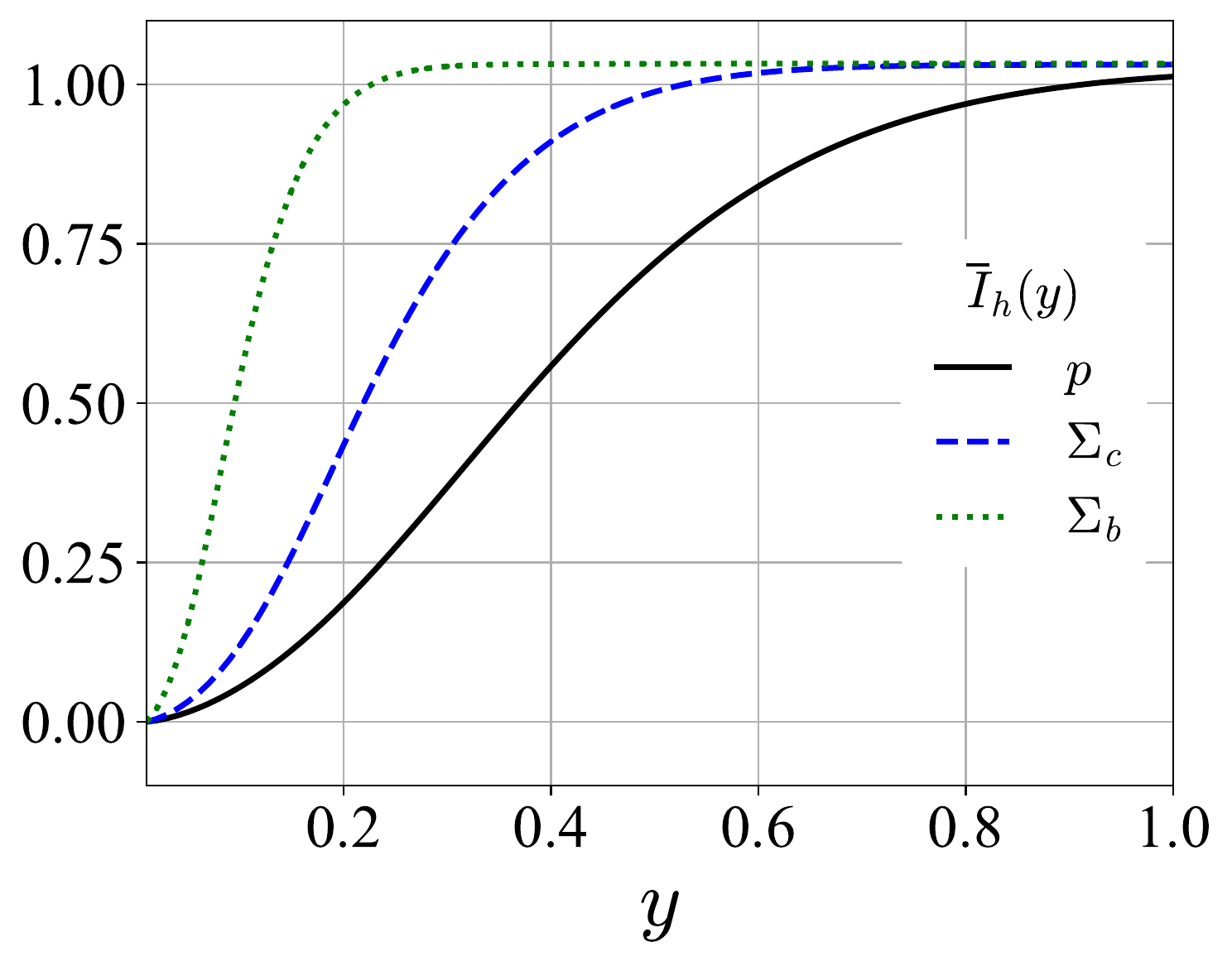}
    }
    \caption{In the left panel, we draw the light-quark momentum
      integral defined in Eq.~\eqref{eq:sum_int_I} for the proton
      ($p$, black solid curve), $\Sigma_c$ (blue dashed one), and
      $\Sigma_b$ (green dotted one). The right panel depicts the
      same curves but normalized by their maximum values, $\bar
      {I_h}(y):= I_h(y) / I_h(y=1)$.} 
    \label{fig:3}
\end{figure}
In Fig.~\ref{fig:3}, the momentum sum given by
Eq.~\eqref{eq:sum_int_I} is plotted for the proton (solid curve), 
$\Sigma_c$ (dashed one), and $\Sigma_b$ (dotted one). 
We observe that the definite momentum shared by the light quarks
inside a singly heavy baryon is smaller than that in a nucleon. This
is due to the fact that the light-quark soliton mass is smaller
in the case of the singly heavy baryons. The light quarks are less
energetic in a singly heavy baryon, compared with those in a nucleon.  
Changing the heavy quark mass does not affect the value of this sum
$I_h(y=1)$,  as far as the $1/M_Q$ heavy-light quark interactions are
not taken into account. However, varying the heavy quark mass
influences the size and the position of the quark and antiquark
distributions. For instance, we find that the $y$ value satisfies
$\bar I_\Sigma(y)=0.8$ are $y=(0.35,0.15)$ for the heavy quark masses
$M_Q=(1.3,4.2)~$GeV, as demonstrated in the left panel of
Fig.~\ref{fig:3}. The right panel of Fig.~\ref{fig:3} shows the 
normalized momentum sum: $\bar {I_h}(y):= I_h(y) / I_h(y=1)$. 

Let us consider the case where the characteristic size $R$ of the pion
mean-field shrinks, i.e., $R \to 0$. In this limit, the $\chi$QSM is
reduced to the naive quark-model, where the nucleon contains only the
non-interacting $N_c$ constituent quarks. For the nucleon, the
isoscalar unpolarized distribution function becomes a delta function
as follows:  
\begin{align}\label{eq:proton_isu_naive}
    u(x)+d(x) = N_c \delta(x-m/M_N),
\end{align}
where $m$ designates the constituent quark mass and $M_N = N_c m$. 
When it comes to the heavy baryon, on the other hand, we obtain
\begin{align}\label{eq:heavy_isu_naive}
    u(x)+d(x) &= (N_c-1) \delta(x-m/M_h), \cr 
    c(x) &= \delta(x-M_Q/M_h).
\end{align}
The singly heavy baryon mass is identified as $M_h=(N_c-1)m + M_Q$.
The momentum sum rule then reads
\begin{align}
    \int^1_0 dx\;x\;[u(x)+d(x) +c(x)] = (N_c-1) m/M_h + M_Q/M_h = 1.
\end{align}

From Eqs.~\cref{eq:proton_isu_naive,eq:heavy_isu_naive},
we observe clearly that the center of the valence-like light-quark PDFs 
is shifted to smaller $x$: from $1/N_c$ to $1/(N_c+M_Q/m)$. 
In a more realistic case, where the heavy and light quark interaction of 
order $1/M_Q$ is included, the heavy quark distribution has a non-zero 
width, as demonstrated in the heavy-quark -- diquark picture 
of the singly heavy baryons in Ref. \cite{Guo:2001wi}. In the current
work, such an effect is ignored as mentioned earlier. 

Last but not least, we want to emphasize that the singly heavy baryon
in the current picture is subject to a certain hierarchy arising from
the parametlically large $M_Q$ and $N_c$. We have constructed the
model first by taking the limit of $M_Q \to \infty$ and then of $N_c
\to \infty$.  This results in the following hierarchy for various
scales:   
\begin{equation}\label{eq:hierearchy}
  1 < M/\Lambda_{\mathrm{QCD}} \ll N_c < M_Q /\Lambda_{\mathrm{QCD}} ,
\end{equation}
where $M$ denotes the dynamical quark mass and
$\Lambda_{\mathrm{QCD}}$ is called the QCD scale parameter. Note that
the parameters for the instanton vacuum are related to
$\Lambda_{\mathrm{QCD}}$~\cite{Diakonov:1987ty, Diakonov:1995qy}.   
In this scheme, the mass of the singly heavy baryon is given by
$M_h=M_{\mathrm{sol}} + M_Q$, where $M_{\mathrm{sol}}$ is proportional
to $N_c$. The momentum sum rule of the light baryon
\cref{eq:momentum_sumrule} vanishes as $N_c \Lambda_{\mathrm{QCD}}/M_Q
\to 0$.  This ordering of the hierarchy between the parameters $N_c$
and $M_Q/\Lambda_{\mathrm{QCD}}$ allows one to strip off the heavy
quark from a singly heavy baryon, since the heavy quark mainly remains
as a static color source (see Refs. \cite{Kim:2017khv, Kim:2018nqf,
  Kim:2018xlc, Kim:2019rcx, Kim:2019wbg, Kim:2020nug}). 
This ordering is strongly supported by the fact that 
the hyperfine splitting between the heavy baryon states $\sim
\mathcal{O}(1/M_Q)\sim 70\,\mathrm{MeV}$ in the case of the charmed
baryons is much smaller than the strength of the  rotational
excitation energy $\sim \mathcal{O}(N_c)\sim 300\,\mathrm{MeV}$.  
If we reverse the ordering of $M_Q\to\infty$ and $N_c\to\infty$
limits, it is very complicated to consider the singly heavy baryons in
the pion mean-field approach with the heavy-quark flavor-spin symmetry 
taken into account. Having constructed the model, we restore physical
values of $N_c=3$ and $M_Q$. The numerical values for $M$ and $M_Q$
used in the current study satisfies the inequality
\cref{eq:hierearchy}, as they should be. 

\subsection{Inequalities}
\begin{figure}[htbp]
    \centering
    \includegraphics[width=4.9cm]{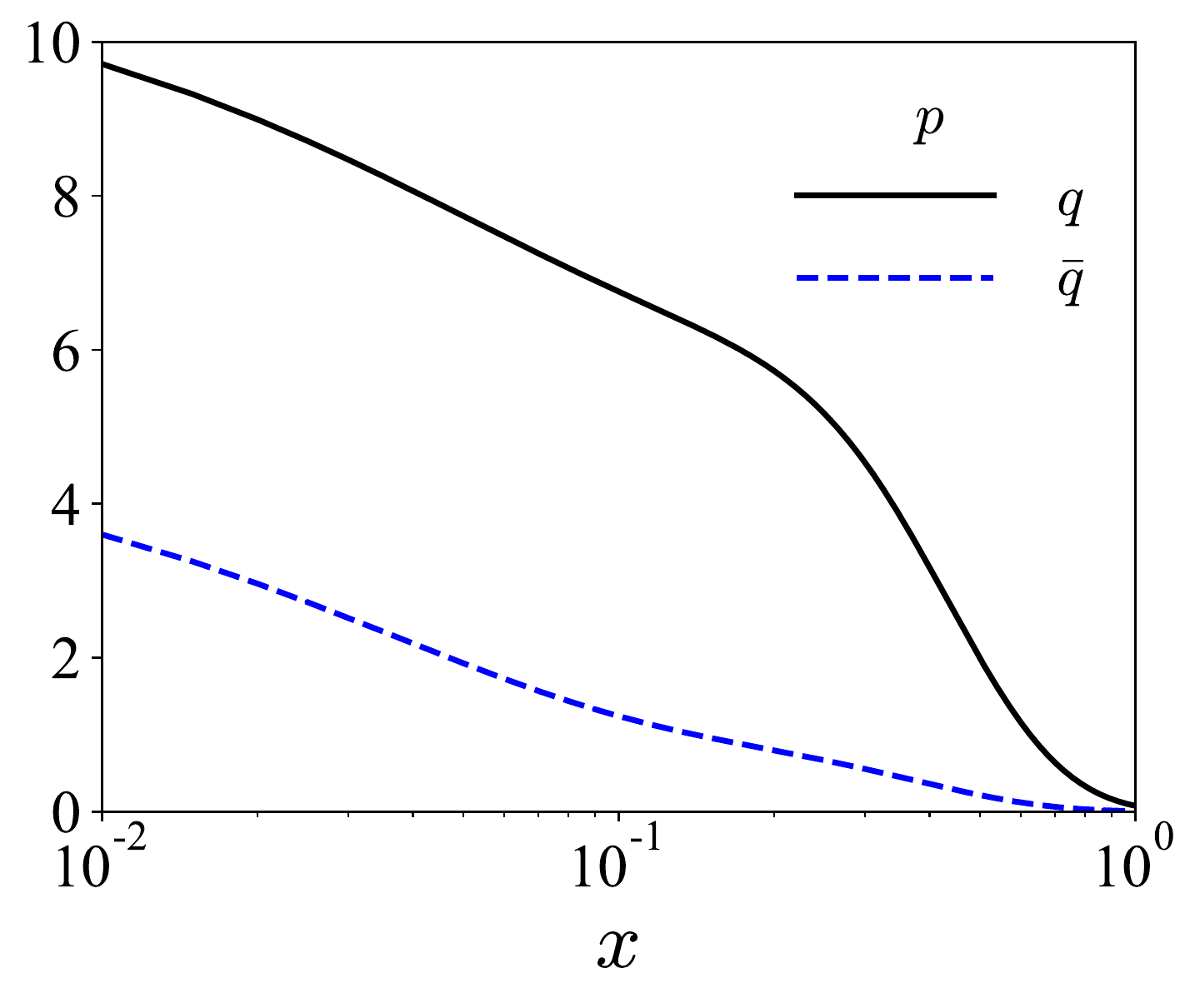}
    \includegraphics[width=4.9cm]{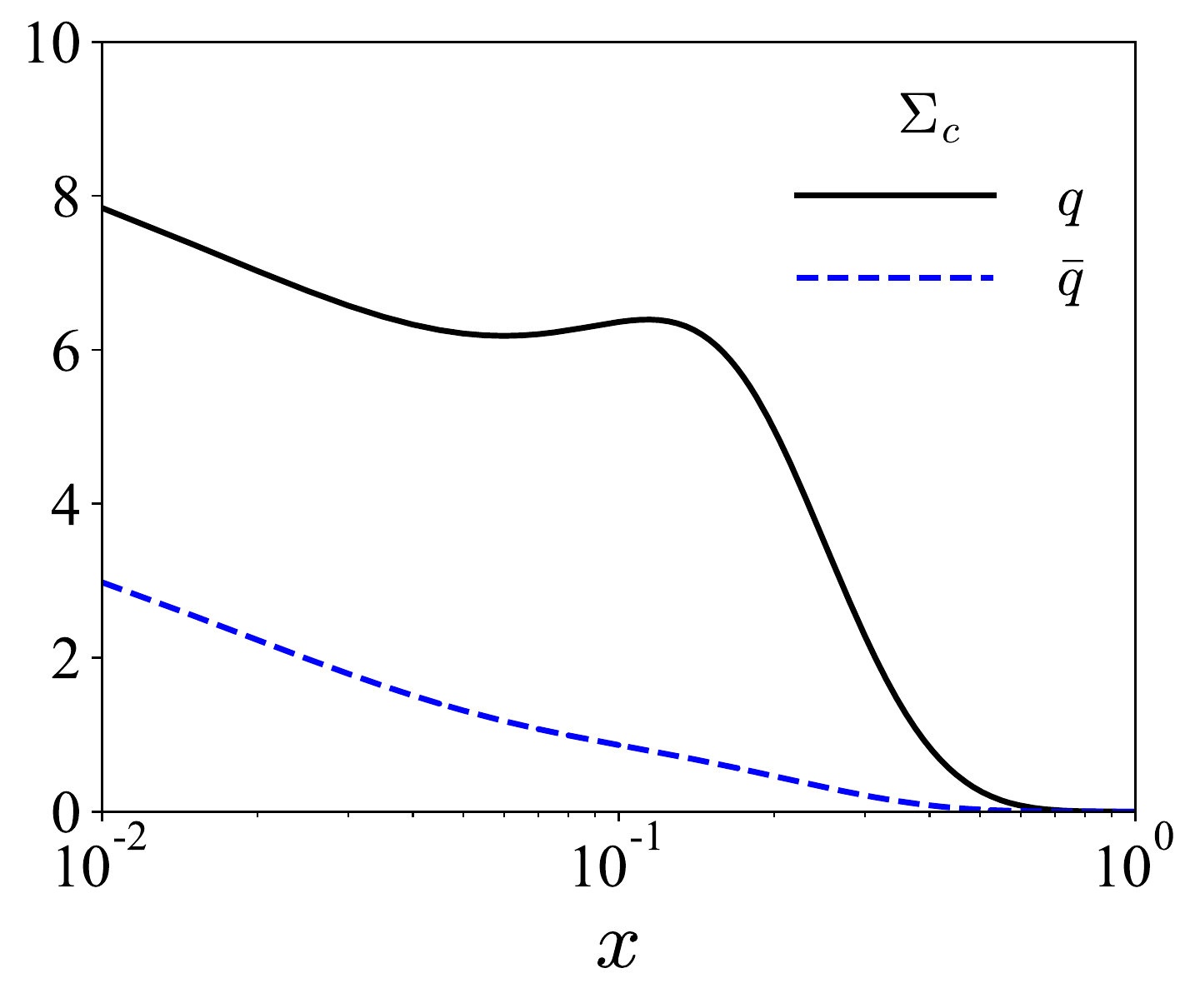}
    \includegraphics[width=4.9cm]{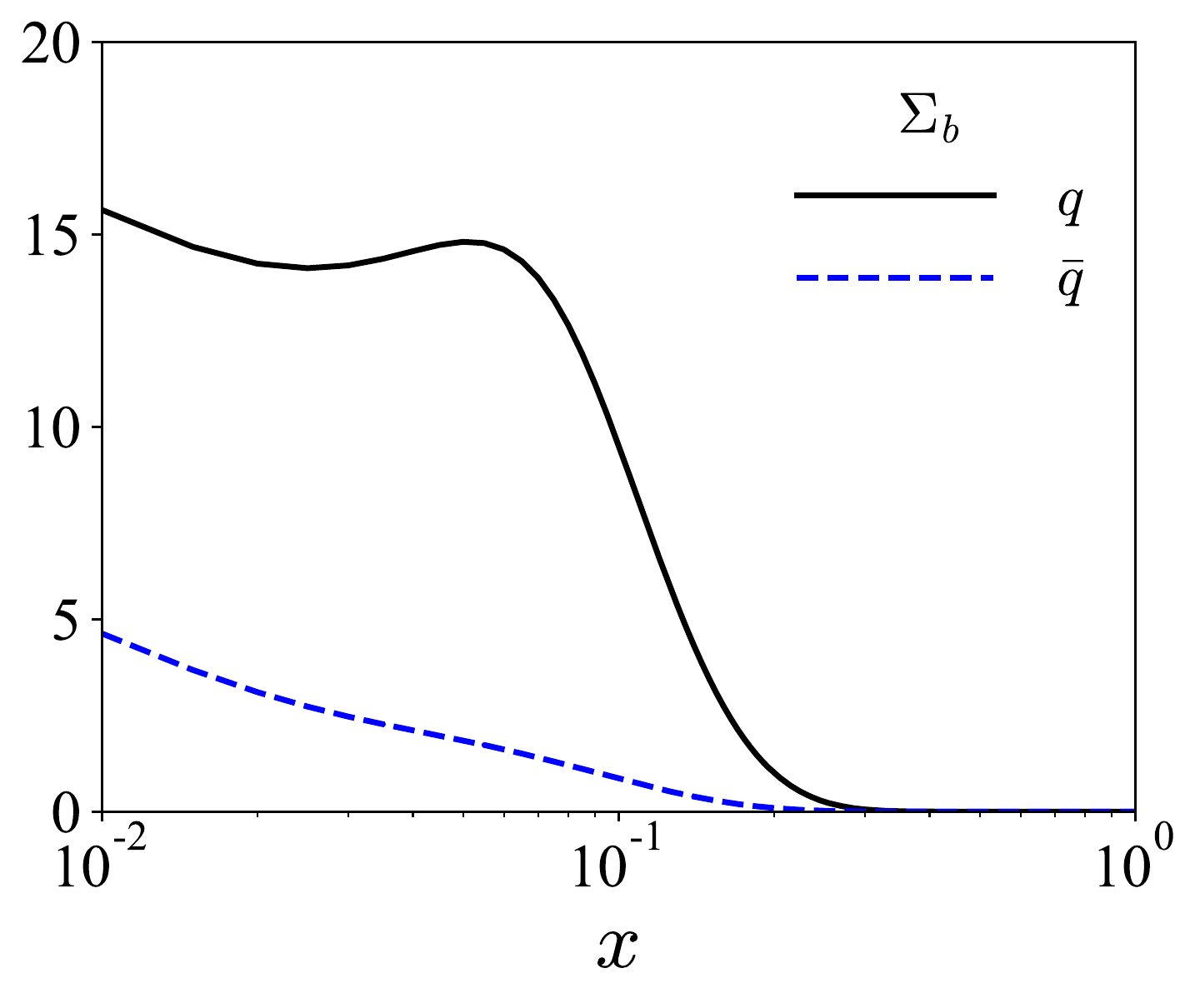}
    \caption{Inequality conditions $ u(x)+d(x) - |\Delta u(x) - \Delta d(x)| $
    for the quark (black solid curve) and antiquark (blue dashed one) are
    plotted for the proton (left panel), $\Sigma_c$ (middle one), and
    $\Sigma_b$ (right one). } 
    \label{fig:4}
\end{figure}

The parton distribution functions are expected to satisfy a set of
inequalities. The inequality conditions related to the present work
are as follows\footnote{Strictly speaking, the positivity condition
  depends on the factorization and regularization schemes. The
  equations shown in the current work are correct to order of
  $\alpha_s$.}:  
\begin{align}
    f^a_1(x) & \ge 0, \label{eq:inequality_f}\\
    f^a_1(x) & \ge |g_1^a(x)| ,
    \label{eq:inequality_f-g}
\end{align} 
where $f_1$ and $g_1$ denote the singlet unpolarized and longitudinally
 polarized parton distribution functions with $a=q, \bar q, g$.
Assumming that the small components $\sim \mathcal{O}(N_c)$ ($u-d$ and
$\Delta u + \Delta d$) vanish in the large $N_c$ limit, we can rewrite
Eqs.~\cref{eq:inequality_f,eq:inequality_f-g} in the following form: 
\begin{align}
    u(x)+d(x) & \ge 0, \label{eq:inequality_f_Nc}\\
    u(x)+d(x) & \ge |\Delta u(x) - \Delta d(x)| \label{eq:inequality_f-g_Nc}.
\end{align}
The positivity and the inequality conditions stem from the probability 
interpretation, so that they provide strong constraints on the PDFs.
Let us examine whether the present results satisfy the inequalities
given by Eqs.~\cref{eq:inequality_f_Nc,eq:inequality_f-g_Nc} in the
 large $N_c$ limit. It was already shown in Fig. \ref{fig:1} that the
 quark and antiquark isoscalar unpolarized distributions satisfy the
 positivity condition Eq.~\eqref{eq:inequality_f_Nc}. 
In Fig.~\ref{fig:4}, the function $ u(x)+d(x) - |\Delta u(x)
- \Delta d(x)| $ is illustrated to check numerically the inequality
Eq. \eqref{eq:inequality_f-g_Nc} for the quarks (solid curve) and the
antiquarks (dashed one). One observes that the inequality condition is
manifestly satisfied for the light and singly heavy baryons within the
$\chi$QSM.

\section{Quark quasi-distribution functions}
\begin{figure}[htbp]
    \centering
    \makebox[\textwidth]{
    \includegraphics[width=6.5cm]{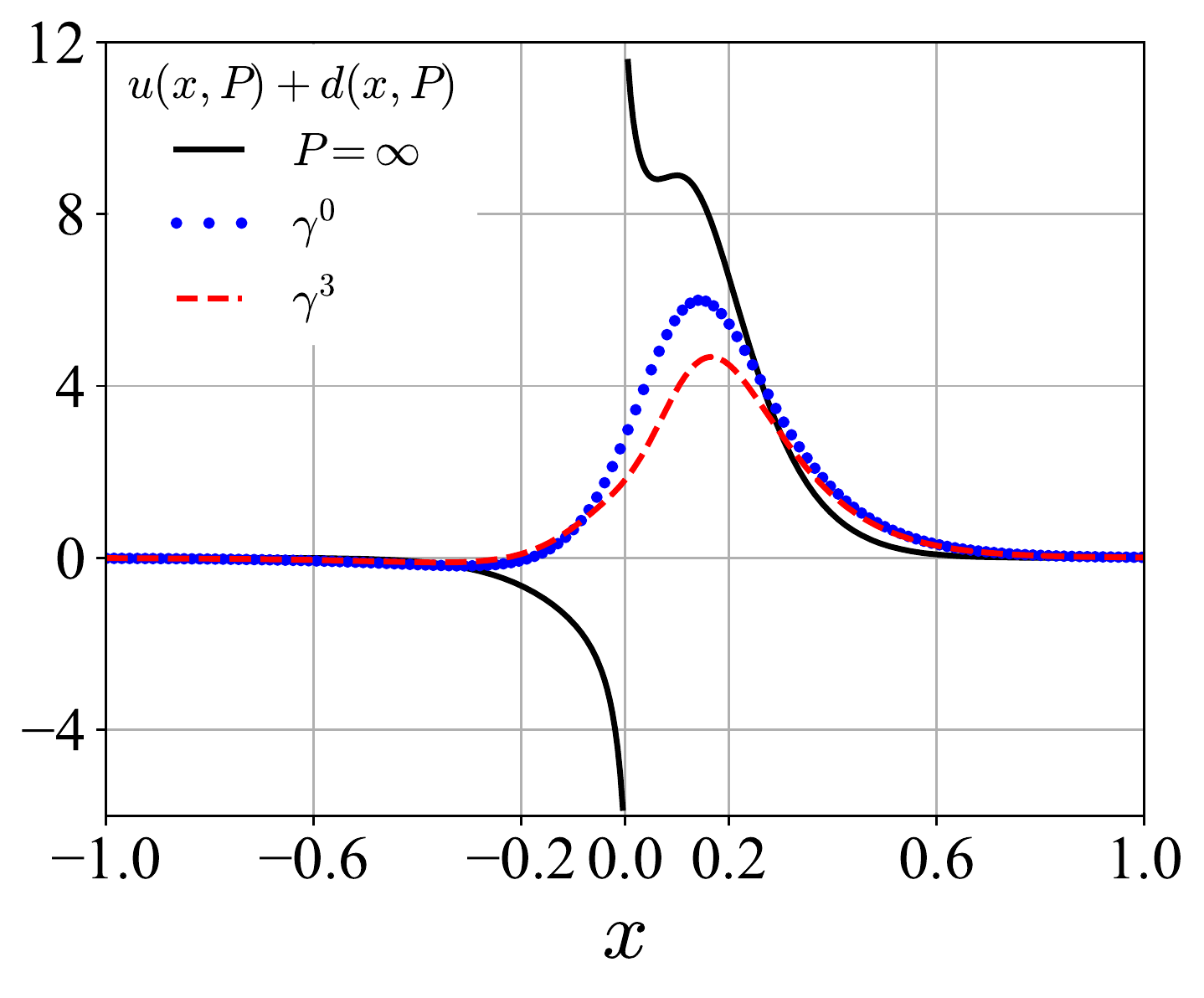}
    \includegraphics[width=6.5cm]{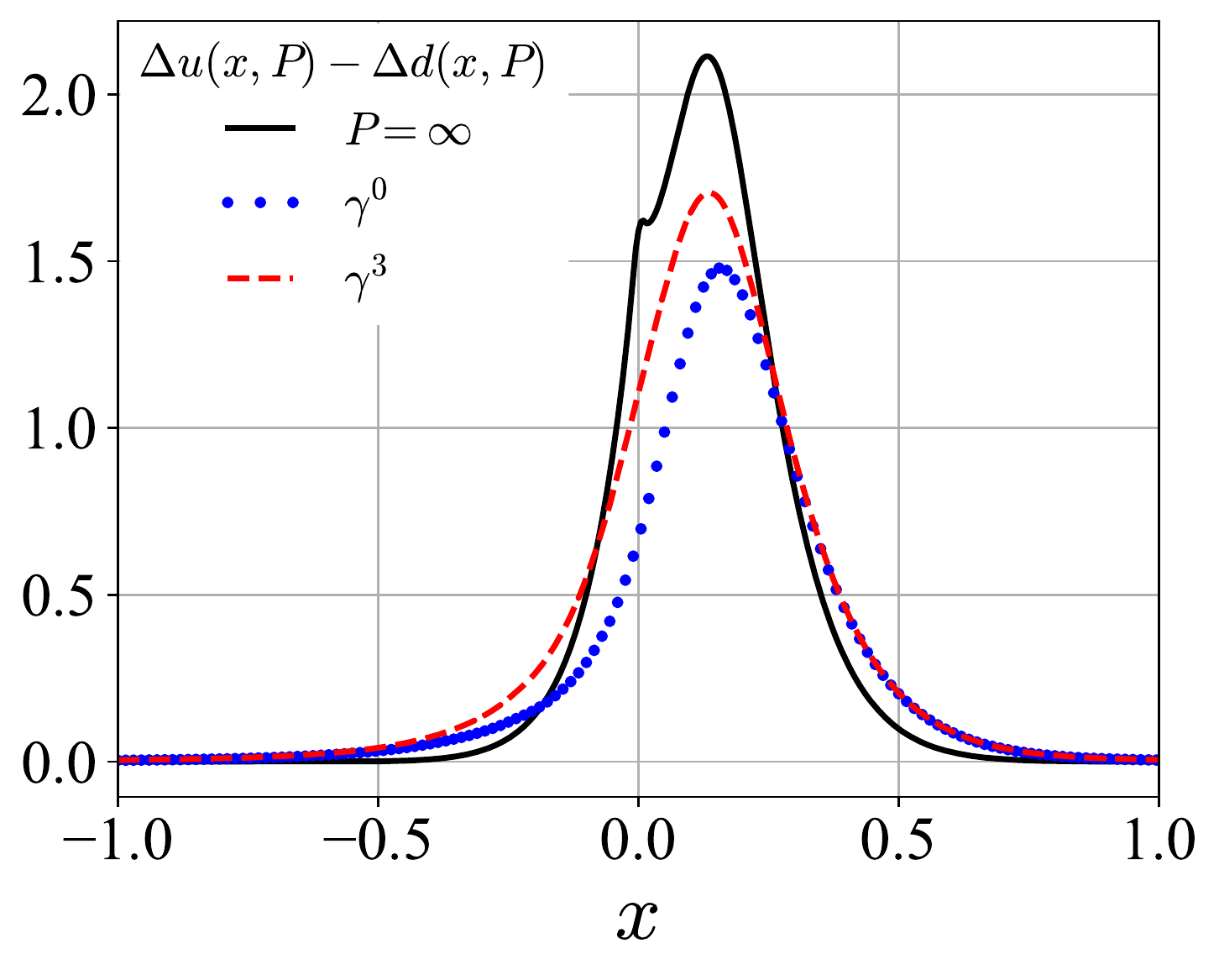}
    }
    \caption{Quark quasi distribution functions in $\Sigma_c^{++}$.
    Isoscalar unpolarized distribution $u+d$ is shown on the left 
    and the isovector polarized distribution $\Delta u - \Delta d$
    is shown on the right panel. Black curves denote the 
    light-cone distribution functions ($P \to \infty$) and 
    the quasi distributions are evaluated with a 
    heavy baryon momentum $P=3~$GeV along the boost direction. 
    Blue dotted and red dashed plots denote $\Gamma=\gamma^0,\;\gamma^3$, 
    respectively.}
    \label{fig:5}
\end{figure}

The PDFs of a singly heavy baryon can also be 
studied in lattice QCD. In recent years, the large momentum effective
theory (LaMET) is widely used~\cite{Ji:2013dva}. In LaMET, one computes the
quasi distributions on the Euclidean lattice and utilizes the pertubative
matching relations to obtain the corresponding light-cone 
distributions. The properties of the
quark quasi-PDFs in the nucleon in the large $N_c$ limit have already
been studied within the $\chi$QSM~\cite{Son:2019ghf,Son:2022qro},
where the convergence of the quasi-PDFs in the nucleon momentum
evolution and the sum rules were discussed. In this section, we
briefly mention about the light-quark quasi distribution functions in
a singly heavy baryon. The model expressions for the twist-2
quasi-PDFs are similar as those in
Refs. \cite{Son:2019ghf,Son:2022qro}. Only differences lie in the number  
of bound-level quarks $N_c-1$ instead of $N_c$ and the corresponding
change of the pion mean field. The Mellin moments of the quasi-PDFs
are related to the physical quantities pertinent to the symmetries,
but they depend on the baryon momentum and the Dirac matrix defining
them~\cite{Son:2019ghf}. In the case of the heavy bayron we have 
the following expressions for the isoscalar unpolarized light-quark
quasi-distributions
\begin{align}
 \int^\infty_{-\infty} \; dx \;( u(x,P)+d(x,P) )&=
 \begin{Bmatrix}
    N_c-N_Q,\;& \Gamma=\gamma^0 \\
   \frac{P}{\sqrt{M^2+P^2}}(N_c-N_Q),\;& \Gamma=\gamma^3
\end{Bmatrix},  \label{eq:quasi_sumrule_bn} \\
 \int^\infty_{-\infty} \; dx \;x( u(x,P)+d(x,P))&= 
  \begin{Bmatrix}
   M_q,\;& \Gamma=\gamma^0 \\
   \frac{P}{\sqrt{M^2+P^2}} M_q,\;& \Gamma=\gamma^3
\end{Bmatrix},
\label{eq:quasi_sumrule_mom}
 \end{align} 
 and for the isovector polarized ones
 \begin{align}
    \int^\infty_{-\infty} \; dx \;( \Delta u(x,P) - \Delta d(x,P) )&=
    \begin{Bmatrix}
    \frac{P}{\sqrt{M^2+P^2}}  (2T_3)\; g_{A,q}^{(3)},\;& \Gamma=\gamma^0 \\
    (2T_3)\; g_{A,q}^{(3)},\;& \Gamma=\gamma^3
   \end{Bmatrix}, 
\label{eq:quasi_sumrule_spin}
    \end{align} 
where $P$ and $M$ denote respectively the heavy baryon momentum and
mass. In Fig.~\ref{fig:5}, the light-quark quasi-distribution 
functions in $\Sigma_c^{++}$ are depicted where the baryon momentum 
$P=3~$GeV is used, which is the typical hadronic momentum 
used in the LaMET framework. The isoscalar unpolarized distribution
$u+d$ is shown in the left panel whereas the isovector polarized
distribution $\Delta u - \Delta d$ is drawn in the right
panel. Especially, one finds that the isoscalar unpolarized
distribution is significantly different from 
the light-cone distribution. The sum rules given in
Eqs. \cref{eq:quasi_sumrule_bn,eq:quasi_sumrule_mom,eq:quasi_sumrule_spin} 
are satisfied numerically.  

Note that the corresponding QCD matrix elements for the quasi-PDFs
can be evaluated on the lattice, in principle. 
However, the numerical results from the current study suggest that 
the LaMET would require unrealistically high momentum for the singly
heavy baryons. A similar observation was found in the case of the heavy
quarkonium distribution amplitude, where the required charmonium
momentum $P$ is 2-3 times its mass~\cite{Jia:2015pxx}.  
Thus, the numerical studies on the Mellin-moment would be more
feasible in a lattice QCD study.  

\section{Summary and conclusions}
In this paper, we studied the properties of the twist-2 light-quark
distribution functions in the large $N_c$ limit inside a singly heavy
baryon within the frame work of the chiral quark-soliton model.
In the limit of the infinitely heavy-quark mass, 
the heavy quark remains as a mere static color source, which provides
heavy quark spin-flavor symmetry. This indicates that the light quarks
govern the quark dynamics inside a singly heavy baryon. Thus, we
focussed on the isoscalar unpolarized $u+d$ and isovector
longitudinally polarized $\Delta u - \Delta d$ quark distributions. 

In the limit of $M_Q \to \infty$, the light quark distributions would
become a $\delta$-function positioned at $x=0$, whereas the heavy
quark carries the entire momentum of the baryon. In reality, however,
the heavy-quark mass is large but finite, so the light-quark  
distributions have finite size and are centered roughly at $x\equiv
M/M_h$ For the numerical calculation, we took heavy-quark masses 
$M_c=1.3$~GeV and $M_b=4.2$~GeV to demonstrate the light-quark  
distributions inside $\Sigma_c$ and $\Sigma_b$ baryons, respectively.  
Compared to the nucleon case, the light-quark distributions are
squeezed to the small momentum fraction $x$. We numerically comfirm
that the positivity and the inequalities are well satisfied.  

As can be seen from Ref.~\cite{Guo:2001wi}, including the interaction
between the heavy quark and the light quarks mostly affects the shape
and position of the heavy quark distribution in a singly heavy baryon.  
On the other hand, we concentrate on the behavior of the light quarks
inside a singly heavy baryon, assuming that such effects
are only order of $1/M_Q$. The heavy-quark distribution functions can
be investigated within the framework of the $\chi$QSM with the
heavy-light quark interaction considered. This interaction can be
derived from the instanton vacuum and will be discussed in future
studies. 
 
The light-quark distributions in a singly heavy baryon can be studied
also within lattice QCD. Firstly, one can compute their Mellin moments 
to study the momentum balance between the light and heavy quark sector. 
On the other hand, it seems that the current LaMET approach is not 
suitable to study the $x$-dependence of the PDFs, because high
baryon momenta $P_h \gg M_h$ are required. Nevertheless, we briefly
discussed the quasi-PDFs and the extended sum rules as their Mellin
moments as done in Refs.~\cite{Son:2019ghf,Son:2022qro} for the
proton, anticipating future results from lattice QCD. 

\section*{\normalsize \bf Acknowledgments}
We are very grateful to J.-M. Suh for useful discussions about the
axial charges of the singly heavy baryons. This work is supported by
Basic Science Research Program through the National Research
Foundation of Korea funded by the Korean government (Ministry of
Education, Science and Technology, MEST), Grant-No. 2018R1A5A1025563
and 2021R1A2C2093368. 

\bibliography{lqHB_references}

\end{document}